\documentclass[aps,prd,preprint,groupedaddress]{revtex4}
\usepackage{graphicx}
\usepackage{amsmath,amssymb,bm}

\begin{document}


\title{Music of the Spheres:  Teaching Quantum Field Theory at the
Introductory Level}


\author{Ian H.~Redmount}
\email[]{ian.redmount@slu.edu}
\affiliation{Department of Physics\\
Saint Louis University\\
St.~Louis, MO~~63103--2010~~USA}


\date{\today}

\begin{abstract}
Quantum field theory has formed the conceptual framework of most of physics
for more than sixty years.  It incorporates a complete revision of our
conception of the nature of matter and existence itself.  Yet it is rarely
taught, or even mentioned, in introductory physics---from high school, college,
and university survey courses through upper-division ``modern physics'' courses.
This omission is not necessary:  This paper describes an approach through
which the fundamental concepts and consequent insights of quantum field
theory can be grasped, building upon familiar notions from classical and
quantum mechanics.  
\end{abstract}


\maketitle

\section{INTRODUCTION\label{sec1}}

Quantum field theory, or the theory of quantized fields, forms the current
conceptual framework of almost all of physics, and has ever since Richard
Feynman, Julian Schwinger, and Tomonaga Shin-ichiro discovered in~1948 how
to make the theory give finite answers to physical questions.  Only gravitation
lies outside the theory's purview; this is the motivation for searches for
a ``quantum theory of gravity'' or ``Theory of Everything,'' from Schr\"odinger
in the 1940's through today's string theory, M-theory, and loop quantum gravity.
Einsteinian quantum field theory is the foundation of elementary-particle
physics, while the Newtonian limit undergirds our understanding of solid-state
physics, superfluidity, and superconductivity.\footnote{The usual distinction
is between ``relativistic'' and ``nonrelativistic'' theories, but this is
incorrect:  Newtonian mechanics and Schr\"odinger quantum mechanics are just
as relativistic as their Einsteinian and Dirac counterparts, only with a
different group of transformations between reference frames.  I shall
therefore attempt quixotically to rationalize the terminology here.}
The theory affords a sweeping syntheses of many features of physics,
and a profound new understanding of the nature of existence itself.

Given the centrality and scope of the theory, it is remarkably difficult
to find an introductory physics text---whether a high-school text,
an algebra/trig-based college physics text, a calculus-based university
physics text, or even a text for a higher-level ``modern physics''
course---which even mentions quantum field theory, much less explains it.
(Recent texts by Hobson~\cite{hobs2010} and Redmount~\cite{redm2016}
are notable exceptions).  There are semi-popular books which describe it,
but its absence from introductory physics texts and courses is striking.
As a result, for example, intrductory treatments of elementary
particles---which do appear in all the texts---read like classical botany,
all terminology and taxonomy.  It has been proposed to teach string theory
at the high-school level, but this is surely putting the cart before the
horse, as all the motivations for string theory are rooted in issues which
arise in quantum field theory; indeed, string theory \emph{is} a quantum
field theory.  Recently Hobson~\cite{hobs2005,hobs2007,hobs2011,hobs2013}
and Huggins~\cite{hugg2007} have argued forcefully for a more central role
for quantum field theory in the teaching of introductory physics, in
particular to clarify many aspects of quantum mechanics.

Of course it would be difficult even to set up the quantum-field-theoretic
calculation of \emph{anything} at the introductory level.  But the fundamental ideas and breathtaking insights of the theory are readily understandable.
They can be explained by building on concepts from classical and quantum
physics which are accessible in elementary treatments.  My purpose in this
paper is to sketch out one such approach.  The necessary foundations are
described in Secs.~\ref{sec2} and~\ref{sec3}.  The emergence of quantum
field theory from these roots, and the features and implications of the
theory, are detailed in Sec.~\ref{sec4}.

\section{QUANTUM MECHANICS\label{sec2}}

Quantum field theory is the quantum mechanics of fields---extended
systems---e.g., fluids, or electromagnetic fields, or the quantum waves
of elementary particles.  Quantum mechanics is built on two principles,
both inferred from observation in the early decades of the 20th century:
Electromagnetic waves, described classically as continuous fields,
actually come in discrete ``packets,'' ``lumps,'' ``quanta,'' or ``photons.''
And conversely, classical particles of matter, such as electrons, protons,
and neutrons, must be described via continuous wave fields.

The particulate nature---quantization---of electromagnetic waves was
introduced by Max Planck in~1900, in order to get the spectrum of radiation
from hot bodies right.  (This is usually considered the opening shot of the
quantum revolution; classical electromagnetic theory fails miserably here.)
Albert Einstein invoked it in~1905 to explain the photoelectric effect,
and Arthur Holly Compton demonstrated it with his x-ray scattering experiments
in~1923.  The mechanical properties of the particles or photons, viz.,
energy~$E$ and momentum~$p$, are related to the properties of the
corresponding waves, viz., frequency~$\nu$, angular
frequency~$\omega=2\pi\nu$, wavelength~$\lambda$, and
wave number~$k=2\pi/\lambda$, via the familiar relations
\begin{subequations}
\label{eq01}
\begin{equation}
\label{eq01a}
E=h\nu=\hbar\omega\ ,
\end{equation}
and
\begin{equation}
\label{eq01b}
p=\dfrac{h}{\lambda}=\hbar k\ ,
\end{equation}
\end{subequations}
where $h=2\pi\hbar=6.626\ldots\times10^{-34}~\hbox{J s}$ is the ``fundamental
quantum of action'' introduced by Planck.  These equations are not as radical
as they might appear:  They accord with the classical relations for the
energies and momenta of electromagnetic wave packets which follow from
the Maxwell equations. They do not, however, represent a throwback to
Newton's corpuscular theory of light. The phenomena which reveal the wave
nature of light---interference and diffraction---remain, even when the
experiments are performed with one photon at a time.  Wave and particle
natures coexist, the famous wave-particle duality of quantum mechanics.

The converse principle, that matter particles exhibit behaviors which must be
described with waves, was revealed during the same period by experiments akin
to those that originally revealed the wave nature of light.  The scattering
of electrons and neutrons from crystals, e.g., in the experiments of Max von
Laue and Davisson and Germer, displayed interference patterns.  Schr\"odinger
tunneling of particles through potential barriers they do not have the energy
to surmount is simply the counterpart of a phenomenon in wave optics, where
it is known as Frustrated Total Internal Reflection.  The relationships
connecting matter-wave properties with particle properties were formulated
by Louis de~Broglie in~1921:
\begin{subequations}
\label{eq02}
\begin{equation}
\label{eq02a}
\lambda=\dfrac{h}{p}\qquad\hbox{i.e.,}\qquad k=\dfrac{p}{\hbar}
\end{equation}
and
\begin{equation}
\label{eq02b}
\nu=\dfrac{E}{h}\qquad\hbox{i.e.,}\qquad \omega=\dfrac{E}{\hbar}\ ,
\end{equation}
\end{subequations}
analogous to the photon relations above.

Incorporating the wave nature of particles into mechanics fundamentally alters
the nature of the science; the questions which are asked and answered are
changed.  A basic problem in classical mechanics is to determine the
trajectory~$x(t)$ of a particle, given its initial position and momentum~$x_0$ and~$p_0$. A wave description does not allow this---that is the content of the
Heisenberg Uncertainty Principle.  In quantum mechanics a basic problem is
the determination of a particle's wave function~$\psi(x,t)$, given its initial
configuration~$\psi(x,0)$.  The squared magnitude of the wave function is
associated with the probability density for finding the particle, in the
standard or Copenhagen interpretation of the theory.  Alternative interpretations have been and continue to be proposed and debated, but all
are \emph{interpretations.}  They change the outcome of no calculation or
prediction of the theory.

The basic ``equation of motion'' in ordinary quantum mechanics, the
Schr\"odinger wave equation, is obtained from the Newtonian energy-momentum
relation
\begin{subequations}
\label{eq03}
\begin{equation}
\label{eq03a}
\dfrac{p^2}{2m}+U=E\ ,
\end{equation}
for a particle of mass~$m$ with potential energy~$U$, by imposing the
de~Broglie relations~\eqref{eq02} in the form of differential operators
applied to the wave function.  Calculus speakers will recognize the
resulting form:
\begin{equation}
\label{eq03b}
-\dfrac{\hbar^2}{2m}\,\nabla^2\psi+U\,\psi=
i\hbar\dfrac{\partial\psi}{\partial t}\ ,
\end{equation}
with $i=\sqrt{-1}$ the imaginary unit, and the time-independent form
\begin{equation}
\label{eq03c}
-\dfrac{\hbar^2}{2m}\,\nabla^2\psi+U\,\psi=E\,\psi\ ,
\end{equation}
\end{subequations}
for wave functions describing particle states of definite energy~$E$.  Despite
the formidable appearance of these equations, their physical content is simply
that of the energy-momentum relation~\eqref{eq03a}.

There are only five exactly, analytically soluble problems in quantum
mechanics: the free particle, the ``particle in a box,'' the free rotor, the hydrogen atom, and the harmonic oscillator.  (If this seems rather limited,
it may be recalled that there are not that many more exactly, analytically
soluble problems in \emph{classical} mechanics.)  The last of these is the
key to understanding quantum field theory.  For an oscillator in one
dimension, with mass~$m$ and ``spring constant''~$k=m\omega^2$, the
time-independent Schr\"odinger equation~\eqref{eq03c} for the energy
levels of the system takes the form
\begin{subequations}
\label{eq04}
\begin{equation}
\label{eq04a}
-\dfrac{\hbar^2}{2m}\,\dfrac{d^\psi}{dx^2}+\tfrac{1}{2}m\omega^2\,\psi
=E\,\psi\ .
\end{equation}
The resulting wave functions are
\begin{equation}
\label{eq04b}
\psi_n(x)=\left(\dfrac{m\omega}{\hbar}\right)^{1/4}(2^n n!\sqrt{\pi})^{-1/2}
H_n\left(\sqrt{\dfrac{m\omega}{\hbar}}\,x\right)\,e^{-\tfrac{1}{2}
\dfrac{m\omega}{\hbar}\,x^2}\ ,
\end{equation}
with $H_n$ the Hermite polynomials discovered in the 19th century.  The
corresponding energy levels are
\begin{equation}
\label{eq04c}
E_n=(n+\tfrac{1}{2})\,\hbar\omega\ ,
\end{equation}
where $n$ is a non-negative integer.  The oscillator has
\emph{zero-point energy}
\begin{equation}
\label{eq04d}
E_0=\tfrac{1}{2}\hbar\omega
\end{equation}
\end{subequations}
in its lowest-energy or ground state.  And the excitation energies of all the
levels above this come in equal increments $\Delta E=\hbar\omega$.  These
features will prove crucial in what is to follow.

\section{DYNAMICS OF SYSTEMS:  NORMAL MODES\label{sec3}}

The other physical principle underlying quantum field theory is a result
from classical mechanics:  The dynamics of any system, however complex,
obeying a linear equation of motion---including almost any system slightly
perturbed from any equilibrium---can be described via a collection of
independent harmonic oscillators.  The most general motion of the system
can be represented as a combination of patterns called \emph{normal modes}
or \emph{harmonics}.  (The harmonics of the vibrating strings or air columns
of musical instruments are normal modes.)  And each normal mode behaves as an
individual harmonic oscillator, uncoupled from the others.

\begin{figure}
\includegraphics{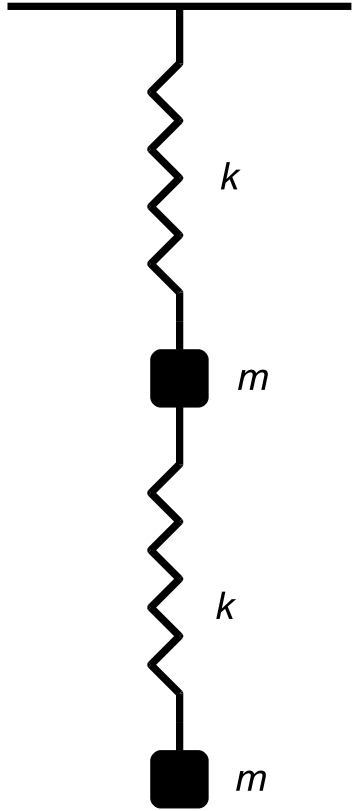}
\caption{\label{f01}The not-quite-simplest system of two coupled harmonic
oscillators.}
\end{figure}

Consider, for example, the system illustrated in Fig.~\ref{f01}:  one simple
harmonic oscillator hung from another, one of the simplest coupled systems
imaginable.  Even if the oscillators have equal masses~$m$ and identical
spring constants~$k$, the springs are essentially massless, and friction and
drag are negligible, the motions of the masses about their equilibrium
positions are complicated and (ideally) nonperiodic.  Yet they can always be
represented as a superposition of simple harmonic motions with two different
frequencies:
\begin{subequations}
\label{eq05}
\begin{equation}
\label{eq05a}
\begin{aligned}
y_1(t)&=c_-\,\cos\left(\dfrac{\omega}{\phi}\,t-\delta_-\right)
+c_+\phi\,\cos(\phi\omega t-\delta_+)\\
y_2(t)&=c_-\phi\,\cos\left(\dfrac{\omega}{\phi}\,t-\delta_-\right)
-c_+\,\cos(\phi\omega t-\delta_+)\ ,
\end{aligned}
\end{equation}
with $c+\pm$ the amplitudes and $\delta_\pm$ the phase shifts of the
normal-mode motions, $\omega=(k/m)^{1/2}$ the angular frequency of a
single oscillator, and
\begin{equation}
\label{eq05b}
\phi=\dfrac{1+\sqrt{5}}{2}=1.618\ldots
\end{equation}
\end{subequations}
the Golden Ratio.\footnote{That the Golden Ratio of the ancient Greeks should
appear in this simple but nontrivial mechanics problem is one of nature's
undeserved gifts to humanity.}  Each of the two terms on the right-hand sides
of Eqs.~\eqref{eq05a} represents the contribution of one of the system's
normal modes to the motion of each mass.  For any system, the number of
independent normal modes is equal to the number of dynamical degrees of
freedom the system possesses.

\section{QUANTUM FIELD THEORY\label{sec4}}

The emergence of quantum field theory from quantum mechanics was impelled,
first, by the need to apply quantum-mechanical principles to electromagnetic
fields, and second, by the need to shift quantum mechanics from its foundation
in Newtonian mechanics to a foundation in Einsteinian mechanics, i.e., in the
dynamics associated with the Special Theory of Relativity formulated by
Albert Einstein in~1905.  In fact these impeti are the same, as classical
electromagnetic fields are inherently Einsteinian.  The bad news is that this
shift is extremely difficult:  The conceptual frameworks of quantum mechanics
and Einsteinian mechanics are almost inconsistent with one another.  The good
news is that if one finds a way to do this, it's probably correct:  There
cannot be too many competing options.

\subsection{Free fields and particles\label{sec4a}}

\subsubsection{Einsteinian Quantum Mechanics\label{sec4a1}}

It is straightforward to construct a quantum-mechanical wave equation
by applying the de~Broglie relations~\eqref{eq02} to the Einsteinian
energy-momentum relation for a free particle of mass~$m$:
\begin{subequations}
\label{eq06}
\begin{equation}
\label{eq06a}
\dfrac{E^2}{c^2}-p^2=m^2c^2\ ,
\end{equation}
with $c$ the vacuum speed of light.  The result is the Klein-Gordon equation,
introduced by Oskar Klein and Walter Gordon in~1926:
\begin{equation}
\label{eq06b}
\hbar^2\left(\dfrac{1}{c^2}\,\dfrac{\partial^2}{\partial t^2}-\nabla^2\right)
\,\varphi+m^2c^2\,\varphi=0\ ,
\end{equation}
\end{subequations}
where I follow convention in using $\varphi$ for the wave function in place
of~$\psi$.  This should serve as the Einsteinian version of the Schr\"odinger
equation~\eqref{eq03b}.  The physics it describes, however, is distinguished by
certain curious features:  Unlike the Schr\"odinger equation, the Klein-Gordon
equation does not define a conserved, positive-definite probability density
(for which the total probability of finding a particle is constant in time).
And the solutions of the Klein-Gordon equation are equally divided between
positive-energy and negative-energy states.  These would appear to be
fundamental flaws in the theory.

In part to circumvent these flaws, Paul Adrien Maurice Dirac introduced
in~1927 the equation which bears his name:
\begin{equation}
\label{eq07}
(\gamma^\mu\,\hat{p}_\mu-mc)\,\Psi=0\ ,
\end{equation}
where $\hat{p}_\mu$ is a differential operator in accord with the de~Broglie
relations, and the first term on the left is summed over the four dimensions
of spacetime.  The price paid for the apparent simplicity of this equation
is that this is a \emph{matrix} equation:  The $\gamma^\mu$ are
$4\times4$~matrices.  The wave function~$\Psi$ is not a single function,
but is represented by a $4\times1$~column matrix; it is an object called a
\emph{Dirac spinor.}

The Dirac equation---with electromagnetic forces incorporated---is chemists'
go-to equation for calculating the energy levels of electrons in atoms and
molecules correctly.  It reproduces the successes of Schr\"odinger quantum
mechanics, but includes corrections associated with Einsteinian mechanics.
Moreover, it describes a \emph{spinning} electron, with intrinsic angular
momentum $\hbar/2$ but corresponding magnetic dipole moment \emph{twice} that
which would be expected from a classical spinning, charged particle.  It is
possible to describe a spinning particle in the context of Newtonian, i.e.,
Schr\"odinger quantum mechanics, but the spin and its interactions must be
put in ``by hand.''  In Dirac's formulation, these features are built
in---there is no non-spinning Dirac particle.  And that factor of~2 in
the magnetic dipole moment---called the ``reduced gyromagnetic ratio''
of the electron---is \emph{almost} right.  Getting it \emph{exactly} right
is one of the triumphs of the quantum-field-theoretic approach.

Dirac's approach does not dispense with the positive and negative energies
of the Klein-Gordon treatment.  To reconcile this with observation---the
electrons around us do not fall through an endless sequence of negative-energy
states---Dirac posited a ``sea'' of electrons which fill all the
negative-energy states.  These are not ordinarily observed, because they
fill space uniformly.  But if sufficient energy is supplied, an electron
can be promoted from the negative-energy sea into a positive-energy state,
where it can be observed.  The vacancy or ``hole'' created in the sea behaves
exactly like a positive-energy electron with opposite electric charge---an
``antielectron'' or ``positron.''  This is not science fiction:  The positron
was discovered by Carl Anderson in~1932.  The antiproton and antineutron were
observed later.  All known particles have corresponding antiparticles.

The Dirac formulation of Einsteinian quantum mechanics thus yields some
spectacularly successful predictions.  But it actually abandons the notion
of a single particle, introducing a multi-particle description even of an
isolated electron.  This calls for the more general framework of quantum
field theory.

\subsubsection{``Second quantization''\label{sec4a2}}

The desired multi-particle, Einsteinian quantum theory is obtained by
applying the principles of quantum mechanics to fields, entities defined
throughout some region of space:  electromagnetic fields, or the quantum-wave
fields which describe other particles.  The application of quantum dynamics
to wave fields from quantum mechanics is described by the historic term
``second quantization,'' although no quantity is ``quantized''
twice.\footnote{Ordinary quantum mechanics is ``first quantization.''
There is such a thing as ``third quantization''; it's an attempt to
incorporate gravity and spacetime geometry into the picture.  The states
of the theory are characterized by different numbers of universes.  The
ground state of the theory is the ``no-universe state,'' though the physical
meaning of that is not entirely clear.}  The values of the field throughout
space at one time become the dynamical variables of the theory, like the
coordinates of a particle in ordinary quantum mechanics.  The fundamental
object of the (second-) quantized theory is the ``wave
functional''~$\Psi[\psi(x),t]$, which assigns a probability amplitude to
the field configuration~$\psi(x)$ at time~$t$.  The squared magnitude
of~$\Psi$ is the probability density for finding the field in that
configuration.  The wave functional (so called because its argument
is an entire function, not simply a number) is the complete description
of the quantum state of the field.  For technical reasons, most calculations
in quantum field theory do not use the wave functional, but an operator
formalism like that of Heisenberg.  There are problems, however, for which
the ``functional Schr\"odinger'' formalism is useful.

This is where the basic principles of the theory come together:  The
field~$\psi$ can be decomposed into its harmonics or normal modes, each
of which behaves as an independent harmonic oscillator.  The quantum field
theory, then, describes a collection of quantum harmonic oscillators, one
for each normal mode.  Each has a zero-point energy and excitations in
uniform energy increments, as in Eqs.~\eqref{eq04c} and~\eqref{eq04d}.
\emph{These energy increments are the particles of the theory;} the numbers
of excitations in all the normal modes define the state of the field.

The description of the particles of matter not as ``lumps of stuff,'' but
as excitations of field modes---more like musical notes than classical
particles---is a sea change in our basic description of existence itself.
We do not \emph{hear} the ancient Greeks' ``Music of the Spheres,'' nor
do we make it:  We \emph{are} the Music of the Spheres.

The payoffs, in terms of understanding, of this approach are huge.  The states
of the theory are defined by numbers of particles in each mode.  The quantum
transitions of the theory, then, can involve both changes of state (mode) of
a particle, or changes in the number of particles.  In any other formulation
of physics---Newtonian, Einsteinian (special- or general-relativistic),
ordinary quantum-mechnical---particle number is a conserved quantity.  But
nature does display processes in which particles are created or destroyed:
An electron and a positron can annihilate, leaving two or three photons; a
collision of protons at Fermilab or the Large Hadron Collider can produce
thousands of new particles.  Only quantum field theory provides a framework
into which such processes fit naturally.

The Dirac equation describes spinning electrons.  In Einsteinian quantum
field theory, all particles are characterized by ``intrinsic spin'' as a
matter of course.  This feature appears as the answer to the question,
``How many different states can describe a single particle?''  Since the
theory is \emph{relativistic,} the answer is ``an infinite number'':  Any
state for a particle, transformed into a different reference frame, must
also be a valid state for that particle.  That is, the particle can appear
with all possible values for its momentum.  Aside from that, then:  ``How
many different states can describe a single, \emph{stationary} particle?''
(This assumes a massive particle.  For massless particles, like the photon,
which are never at rest, the argument differs slightly in detail.)  Since
the theory is quantum-mechanical, the states must form a vector space.  The
\emph{dimension} of the space is tabulated by the spin quantum number of the
particle.  If the space is one-dimensional---there is only one state---the
particle is spinless:  $s=0$.  A two-dimensional space---the states are
combinations of ``spin up'' and ``spin down''---corresponds
to~$s=\tfrac{1}{2}$, as for the electrons, the neutrinos, and the quarks.
Three dimensions mean $s=1$; four, $s=\tfrac{3}{2}$; five, $s=2$; in general
the dimension of the space of stationary-particle states is $2s+1$.  Because
the states can be transformed into one another by all the changes of reference
frame which do not change the velocity of the particle, i.e.,
\emph{rotations,} this tabulator of the number of states takes
on the characteristics of quantum angular momentum.  The intrinsic spin
of particles and all its consequences, such as magnetism, arise in quantum
field theory as a simple bookkeeping device.

An even simpler bookkeeping device has even more profound implications.
Consider a state~$\Psi_{kk^\prime}$ of two particles, the first in mode~$k$
and the second in mode~$k^\prime$.  How is this related to
state~$\Psi_{k^\prime k}$, with the first particle in mode~$k^\prime$ and the
second in mode~$k$?  We observe in nature that elementary particles of the
same species are \emph{absolutely identical;} no individual distinguishing
features have ever been observed.  Classically, this is remarkable, but in
quantum field theory, it is entirely natural, as the particles are excitations
of the same oscillator.  So exchanging the roles of the two particles in
states~$\Psi_{kk^\prime}$ and~$\Psi_{k^\prime k}$ can have no observable
consequences---all the probabilities determined by the two wave functionals
must be the same.  What can you do to a quantum wave functional which
1)~doesn't change its squared magnitude (probability), and 2)~if you do
it twice, doesn't change anything at all, since exchanging the particles twice
undoes the exchange?  There are exactly two possibilities:  Multiply the wave
functional by~$+1$, or multiply it by~$-1$.  Particles for which the former
applies are called ``bosons,'' after Satyendra Nath Bose, and particles for
which the latter applies are called ``fermions,'' after Enrico Fermi.  This
classification is termed the ``statistics'' of the particles or field.
(More precisely, bosons obey ``Bose-Einstein statistics,'' fermions
``Fermi-Dirac statistics.'')  Consider now a state~$\Psi_{kk}$ with two
particles in the \emph{same} mode.  Exchanging the particles here changes
nothing.  But there is only one value which is unchanged upon multiplication
by~$-1$:  Necessarily, for fermions $\Psi_{kk}=0$.  That is, \emph{there is
no probability of finding two fermions in the same mode, or quantum state.}
This is the Pauli Exclusion Principle, introduced by Wolfgang Pauli in~1925.
Like spin, it can be introduced by hand into ordinary quantum mechanics.
But in quantum field theory it is built in, a necessity to keep the
particle-exchange rules consistent. The Pauli Exclusion Principle gives
ordinary matter its rigidity and impenetrability, and undergirds all of
chemistry:  Because electrons in atoms must be stacked at most two to an
orbital (one spin-up, one spin-down), atoms with different atomic numbers
have different electron configurations, hence different chemical properties.
All these aspects of matter, then, are quantum-field-theoretic effects.

It is a remarkable result of Einsteinian quantum field theory that the spin
and statistics characteristics of particles are not independent.  Bosons have
integral spin quantum numbers, fermions half-integral values, and conversely.
This result is known as the Spin-Statistics Theorem.  A proof is beyond the
scope of the present discussion.

\subsection{Interacting fields and renormalization\label{sec4b}}

As important as they are, these features of the quantum theory of \emph{free}
fields are only the kinematics of the theory.  The dynamics of quantum field
theory emerges in the description of interacting fields.  Since a field
entails an infinite number of dynamical variables or ``degrees of
freedom''---e.g., the value of the field at each point in space---such a
description is inherently problematic.  The challenges presented by such
theories have shaped the evolution of physics for the past seventy years.

\begin{figure}
\includegraphics{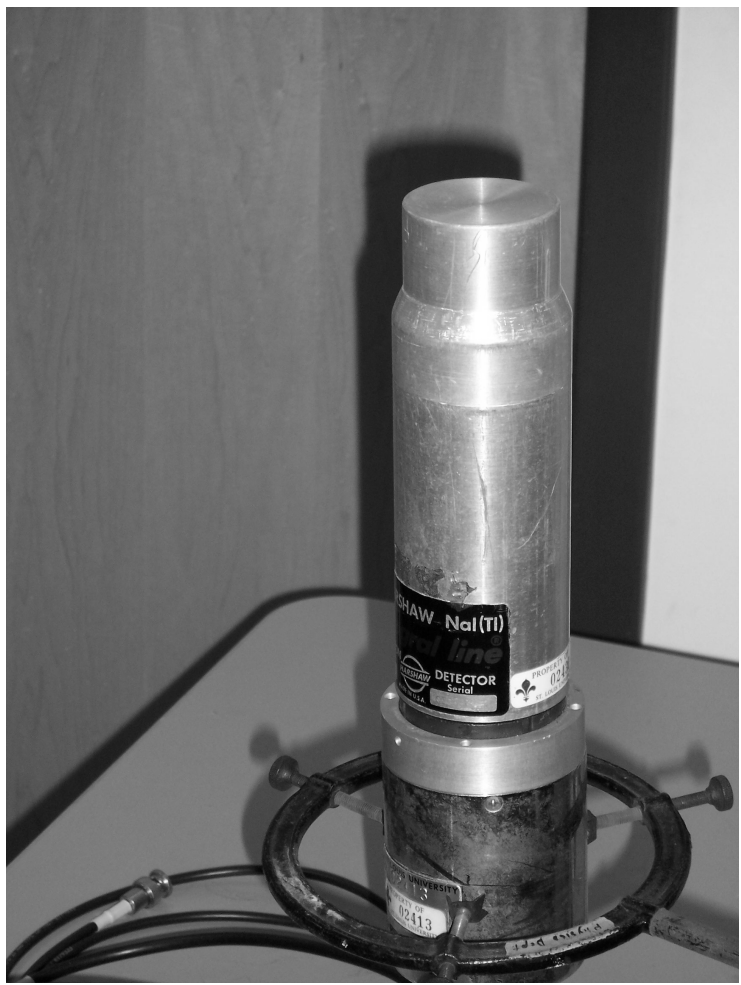}
\caption{\label{f02}A scintillation detector for gamma-ray photons.  Gamma
rays interacting with the NaI crystal (top) produce flashes of visible
light, converted to electrical pulses by the photomultiplier tube
(bottom portion).  This can be understood as a sequence of quantized-field
interactions.}
\end{figure}

It is typical to envision an elementary-particle reaction, such as the decay
\begin{equation}
\mu^-\to e^-+{\bar\nu}_e+\nu_\mu
\end{equation}
of a muon ($\mu^-$) into an electron ($e^-$), an electron antineutrino
(${\bar\nu}_e$), and a muon neutrino ($\nu_\mu$), as an explosion.  But
while this event is far more violent---in terms of the fraction of particle
mass converted into kinetic energy---than any chemical or even nuclear
explosion, such a description is not apt.  The features we associate with
explosions are gas-dynamic phenomena, involving septillions of particles.
The muon decay is closer akin the the production of a note from the strings
of a piano by singing into the piano:  In that case energy is transferred
from the vibrating vocal cords of the singer to the air, then to the sound
board and strings of the piano, then back to the air, then to the ears of
the listener.  In the muon decay, the excitation energy of the muon field
(which constitutes the muon particle) is transfered to the electron,
electron-neutrino, and muon-neutrino fields, appearing as those particles.
Another example is illustrated in Fig.~\ref{f02}, which shows a NaI-crystal
scintillation detector for gamma rays.  Incoming gamma-ray photons produce
electrons in the crystal via Compton scattering, photoelectric effect, or
pair production.  The electrons interact with atoms in the crystal to
produce flashes of visible light, or scintillations.  These interact with
the plates of a photomultiplier tube to produce a cascade of electrons;
the resulting current pulse is detected and analyzed by suitable circuitry.
From a field-theoretic standpoint, the incoming energy is transferred
between the electromagnetic (photon) and electron fields three times.

\begin{figure}
\includegraphics{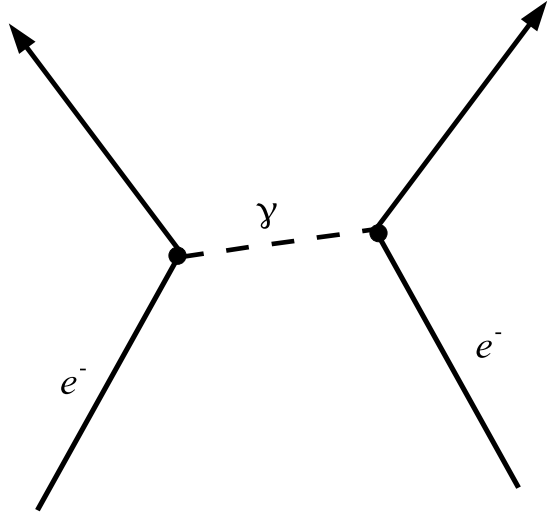}
\caption{\label{f03}The simplest Feynman diagram for the scattering of
two electrons via electromagnetic forces.  This is a representation of
a mathematical expression, not of a sequence of physical events.}
\end{figure}

Because of the complexities involved with infinities of interacting degrees
of freedom, there are no exact analytic solutions of interacting quantum
field theories, except perhaps in a few contrived examples.  Instead, field
interactions are treated as perturbations of free-field theories.  Physical
quantities are calculated via infinite series of ``correction'' terms.  This
approach is represented graphically via the figures known as Feynman diagrams,
introduced by Richard Feynman.  For example, for two electrons scattering via
electrostatic repulsion, the simplest diagram is shown in Fig.~\ref{f03}.
It is tempting to interpret this classically:  The electron ($e^-$) on the
left emits the virtual photon ($\gamma$) and recoils; the electron on the
right absorbs the photon and is deflected.  But how could the attraction of
oppositely charged particles be described in this way?  The first time I
heard this question asked, the answer given was, ``One mustn't take these
things too literally.''  As unsatisfying as that seems, it is in fact the
correct answer.  Feynman diagrams are not literal depictions of events in
physical spacetime; rather, they are mnemonic devices for constructing the
terms in an infinite series for, say, a scattering amplitude or reaction
probability.  The same diagram represents attraction or repulsion; only the
algebraic signs of the corresponding expressions are different.

The perturbation approach has one glaring drawback.  The result of any
calculation of a physical quantity is:  infinity.  The expressions
corresponding to any diagram more complex than Fig.~\ref{f03} are
divergent integrals.  That is, the individual terms of the perturbation
series for any quantity are infinite, not even to speak of the convergence
of the series.  Only after Richard Feynman, Julian Schwinger, and Tomonaga
Shin-ichiro discovered (independently) in~1948 how to evade this quandary
did it become possible to extract physical predictions from quantum field
theory.  The three men shared the Nobel Prize in Physics in~1965 for their
discovery.  It is one of the most fundamental, and at the same time most
obscure, ideas in science.  It is called ``renormalization.''

Renormalization can be described in a variety of ways, not obviously connected
with one another:  It corrects the parameters of the theory for the effects of
the interactions between fields.  It absorbs the inaccessible high-energy
behavior of the theory into its parameters.  It compensates for the fact that
the quantum states of the interacting-field theory do not exist in the same
abstract vector space as the states of the free-field theory.  Perhaps the
simplest way to understand it is to note that any quantum field theory
contains certain parameters, such as the masses and charges of the particles
described, and the scale of the fields.  Given finite values of these
parameters, the theory gives infinite results for any physical quantity.
But by allowing the parameters to become infinite in the opposite direction,
the infinities can be canceled, yielding finite answers to physical questions.

The parameters allowed to become infinite are called the ``bare parameters''
of the theory.  They represent, e.g., the masses and charges of particles,
and the scales of fields, stripped of their interactions.  But no particle
or field is ever observed stripped of its interactions, so there is no
conflict with observation.

The difference between two infinite quantities, i.e., two quantities diverging
to infinity in some suitable limit, can be any finite value.  The finite
values of renormalized quantities are pinned down by requiring them to
match specific meaurements:  For example, the renormalized mass of the
electron at zero interaction energy must match the mass of the electron
measured in low-energy experiments.  This has the consequence that the
parameters of the theory have values which vary with interaction energy;
the theory is said to have ``running coupling constants.''  That looks
like a contradiction, except that nature actually does this:  For example,
the strength of the electric interaction between electrons is measurably
greater at the energies attained in high-energy accelerators than in
low-energy events.  This is one of the signal successes of renormalized
quantum field theory.

The infinities or divergences which arise in field-theory calculations cannot
be dismissed as mere artifacts of the calculation.  They have observable, and
observed, consequences.  The probabilities for some processes consist of a
factor which is formally zero, multiplied by a diverging factor, giving a
finite, nonzero result.  Such processes, called ``anomalies,'' actually occur.

The results of renormalization are spectacularly vindicated by experiment.  Dirac's treatment of the electron predicts that certain states of the
hydrogen atom should be degenerate, i.e., have the same energy.  The
energies actually differ very slightly; the difference can be measured
via microwave spectroscopy.  Renormalized quantum electrodynamics correctly
predicts this difference, known as the ``Lamb shift.''  Dirac's treatment
also predicts that the ratio of the magnetic dipole moment of the electron
to its spin angular momentum---its ``gyromagnetic ratio''---should be exactly
twice that obtained from a classical treatment of the electron as a spinning
ball of electric charge.  The actual factor is about one-tenth of one percent
greater than two.  The discrepancy can be calculated using renormalized
quantum electrodynamics.  At present, calculation and measurement agree to
some eighteen significant figures.  Nowhere else in science does a prediction
agree with observation to such precision and accuracy.\footnote{As the
precision of the calculation is increased, the number of perturbation terms
which must be included grows rapidly, into the hundreds of millions.  This
led in the 1970's and 1980's to an unprecedented situation:  It was more
expensive to \emph{calculate} the next digit in the gyromagnetic ratio than
to \emph{measure} it!}  The precision of this result is now so high that
calculating further digits is beyond the scope of quantum electrodynamics
alone.  The contributions of nuclear interactions must now be included.

As more and more perturbation terms are included in any calculation, more
and more divergences must be absorbed into the same set of parameters.  The
form of the field theory must be rather special for this to be possible.
Such a theory is called ``renormalizable.''  This property,
``renormalizability,'' seems extraordinarily arcane, but it has been
regarded as a \emph{sine qua non} of physical theory for the last seventy
years.  Feynman, Schwinger, and Tomonaga established that quantum
electrodynamics, the theory of electromagnetic fields interacting with
charged particles, is renormalizable.  Gerard t`Hooft and Martin Veltman
were awarded the Nobel Prize in Physics in~1999 for their proof that
Yang-Mills theories---a class of generalizations of quantum
electrodynamics---are also renormalizable.  Our current theories
of both strong and weak nuclear interactions are Yang-Mills theories.
But Einstein's General Theory of Relativity, our best available description
of gravitation, is \emph{not} a renormalizable field theory.  This has
motivated decades of searching for a quantum theory of gravitation,
a so-called ``Theory of Everything.''

Since the 1990's it has been suggested that the ultimate physical theory
might not need to be perturbatively renormalizable.  The hope is that more
sophisticated, nonperturbative calculations---large-scale numerical
calculations, perhaps---might be able to extract the physical content of
the theory, including the effects of its interactions, without generating
and then absorbing infinite terms.  The search is in its early stages, and
has not yet produced definitive results.

\subsection{Newtonian quantum field theory\label{sec4c}}

The conceptual framework of quantum field theory---treating the dynamics of
an extended system via the quantum excitations of its normal modes---is not
restricted to Einsteinian quantum theory and elementary-particle physics.
Quantized fields in the Newtonian limit are employed in other areas.  In
solid-state physics the excitations of the vibrations of an atomic lattice
are called ``phonons,'' quanta of sound or thermal vibration.  They interact
with the electron field in the material.  In nuclear physics, excitations of
the collective motions of protons and/or neutrons in nuclei are called
``quasiparticles.''  Different numbers and states of quasiparticles
characterize different excited states of the nucleus.

\subsection{The Standard Model and beyond\label{sec4d}}

Quantum field theory might be termed a ``meta-theory,'' i.e., a conceptual
framework, within which specific physical theories are formulated by
specifying the fields in play and their interactons.  The current ``Standard
Model'' of particle physics, our best description to date of the fundamental
nature of matter, consists of three Yang-Mills theories combined, treating
the quark and lepton fields which make up the particles we observe, and
their interactions:  the ``color'' interaction between quarks, which
engenders the strong nuclear force (and has nothing to do with optical color);
the weak nuclear forces, and the electromagnetic forces described by
the original quantum electrodynamics of Feynman, Schwinger, and Tomonaga.
This model is consistent with all available data on the physics of elementary
particles.\footnote{Strictly, the Standard Model is a \emph{theory,} i.e.,
a well-tested hypothesis and conceptual framework, while individual Yang-Mills
constructions are \emph{models} or \emph{hypotheses.}  The reverse usage,
however, is traditional.}

A great deal of recent excitement has centered on the 2012 discovery of
the ``Higgs boson'' at the Large Hadron Collider in Geneva, Switzerland.
Higgs bosons are the quantum excitations of the Higgs field.  This field was
proposed in~1964---by Peter Higgs, by Fran\c{c}ois Englert and Robert Brout,
and by Tom Kibble, Carl R.~Hagen, and Gerald Guralnik---in hopes of accounting
for the particle/field masses in the Standard Model as interactions between
fields, rather than as arbitrary parameters of the model.  The identification
of the Higgs particle in data from the Large Hadron Collider has apparently
vindicated these theorists' insights.

Many attempts have been made to merge the three theories making up the
Standard Model into a single Yang-Mills theory, allowing processes in
which quarks, e.g., in nucleons might transform into leptons.  Such
constructs are called Grand Unified Theories (GUTs).  The simplest models,
proposed in the 1970's, made predictions in conflict with later observations
and are thus ruled out.  But efforts to construct a successful Grand Unified
Theory continue today.

Neither the Standard Model nor the Grand Unified Theories incorporate
gravitational interactions between particles.  Our most accurate and
precise understanding of gravitation to date is contained in Einstein's
1915 General Theory of Relativity, which describes gravitation in terms
of the geometry of spacetime.  But this is a classical theory.  Attempts
to construct a quantum theory of gravitation go at least as far back as
Schr\"odinger in the 1940's.  Such attempts are hampered by the fact that
Einstein's theory, expressed as a quantum field theory, is not renormalizable:
Physical predictions cannot be extracted from a quantum version by setting
the values of a finite number of parameters.  Approaches to an eventual
quantum theory of gravitation are varied.  As a first step, ordinary
quantum field theories can be formulated in the curved spacetime of
Einstein's theory---the normal modes of the fields reflect the spacetime
geometry.  Stephen Hawking's 1974 discovery that black holes might
radiate via quantum processes emerged from this approach.  Or Einstein's
theory might be recast as a quantum theory without relying on the usual
perturbative calculations---various attempts in this direction are ongoing.

In the other direction, quantum field theories of the usual form can be
constructed incorporating a gravitational interaction, e.g., via a
``graviton'' field.  Such a model is called a ``super-unified'' theory
or, with some hyperbole, a ``Theory of Everything.'' One class of such
candidate models links particles/fields of different spins (and statistics).
This feature is called ``supersymmetry'' (SUSY) or, when a graviton
field is involved, ``supergravity.''  These have been studied extensively
since the 1970's, but the goal of a satisfactory super-unified theory has
not yet been reached.

Perhaps the most extensively explored extensions of standard quantum field
theory are string theories and a generalization of them, M-theories.  These
are quantum field theories in which the fields ``live'' not on spacetime
points but on extended structures, ``strings'' or ``branes''---short for
``membranes.''  (The M in M-theories stands for ``matrix''; the field values
in these theories are matrices, rather than single real or complex numbers.)
Originating in models for strong nuclear forces proposed in the 1960's, the
current versions of these theories were launched by the work of Michael Green
and John H.~Schwartz in the 1980's.  These theories or, more properly,
hypotheses appear to offer the promise of, first, a finite theory without
infinities; second, a quantum theory incorporating a gravitational
interaction; third, a unique theory---a single, mathematically consistent
theory which must be correct, there being no consistent alternative.
To date, however, these theories have yet to fulfill any of these promises.

\section{SUMMARY\label{sec5}}

Quantum field theory has formed the conceptual framework of much of
contemporary physics since the middle of the 20th century.  It treats the
dynamics of extended systems---from atomic lattices in solids to the fields
which constitute the fundamental structure of matter---in terms of the quantum
excitations of the systems' normal modes.  It can thus describe dynamical
processes in which excitations or particles are created or destroyed.
It allows a consistent melding of quantum and Einsteinian mechanics,
incorporating key features of elementary particles, such as spin and
statistics, in a natural and inevitable way.  It relies on the
renormalization procedure to absorb the infinities or divergences
in a theory into its parameters, but in so doing yields agreements
between calculation and measurement unrivaled anywhere else in science.
Much of the current frontier of theoretical physics lies within the scope
of quantum field theory.

Yet as fundamental and transformative as it is, quantum field theory is
rarely even mentioned in introductory physics, from secondary-school through
upper-division undergraduate courses.  As illustrated here, this omission---%
comparable to omitting DNA from introductory biology---is not necessary.  The
theory can be described using no more advanced concepts than the harmonics
of extended systems and the energy levels of the quantum harmonic oscillator.
It is to be hoped that with such inclusion, the introductory treatment of
physics can be brought into greater harmony with the current state of
the field.

\bibliography{hsqft}

\begin{thebibliography}{7}
\expandafter\ifx\csname natexlab\endcsname\relax\def\natexlab#1{#1}\fi
\expandafter\ifx\csname bibnamefont\endcsname\relax
  \def\bibnamefont#1{#1}\fi
\expandafter\ifx\csname bibfnamefont\endcsname\relax
  \def\bibfnamefont#1{#1}\fi
\expandafter\ifx\csname citenamefont\endcsname\relax
  \def\citenamefont#1{#1}\fi
\expandafter\ifx\csname url\endcsname\relax
  \def\url#1{\texttt{#1}}\fi
\expandafter\ifx\csname urlprefix\endcsname\relax\def\urlprefix{URL }\fi
\providecommand{\bibinfo}[2]{#2}
\providecommand{\eprint}[2][]{\url{#2}}

\bibitem[{\citenamefont{Hobson}(2010)}]{hobs2010}
\bibinfo{author}{\bibfnamefont{A.}~\bibnamefont{Hobson}},
  \emph{\bibinfo{title}{Physics: Concepts \& Connections, 5th Edition}}
  (\bibinfo{publisher}{Addison-Wesley}, \bibinfo{address}{Upper Saddle River,
  NJ}, \bibinfo{year}{2010}).

\bibitem[{\citenamefont{Redmount}(2016)}]{redm2016}
\bibinfo{author}{\bibfnamefont{I.~H.} \bibnamefont{Redmount}},
  \emph{\bibinfo{title}{12 Problems in Physics, First Edition}}
  (\bibinfo{publisher}{Cognella Academic Publishing}, \bibinfo{address}{San
  Diego, CA}, \bibinfo{year}{2016}).

\bibitem[{\citenamefont{Hobson}(2005)}]{hobs2005}
\bibinfo{author}{\bibfnamefont{A.}~\bibnamefont{Hobson}},
  \bibinfo{journal}{American Journal of Physics} \textbf{\bibinfo{volume}{73}},
  \bibinfo{pages}{630} (\bibinfo{year}{2005}).

\bibitem[{\citenamefont{Hobson}(2007)}]{hobs2007}
\bibinfo{author}{\bibfnamefont{A.}~\bibnamefont{Hobson}}, \bibinfo{journal}{The
  Physics Teacher} \textbf{\bibinfo{volume}{45}}, \bibinfo{pages}{96}
  (\bibinfo{year}{2007}).

\bibitem[{\citenamefont{Hobson}(2011)}]{hobs2011}
\bibinfo{author}{\bibfnamefont{A.}~\bibnamefont{Hobson}}, \bibinfo{journal}{The
  Physics Teacher} \textbf{\bibinfo{volume}{49}}, \bibinfo{pages}{12}
  (\bibinfo{year}{2011}).

\bibitem[{\citenamefont{Hobson}(2013)}]{hobs2013}
\bibinfo{author}{\bibfnamefont{A.}~\bibnamefont{Hobson}},
  \bibinfo{journal}{American Journal of Physics} \textbf{\bibinfo{volume}{81}},
  \bibinfo{pages}{211} (\bibinfo{year}{2013}).

\bibitem[{\citenamefont{Huggins}(2007)}]{hugg2007}
\bibinfo{author}{\bibfnamefont{E.}~\bibnamefont{Huggins}},
  \bibinfo{journal}{The Physics Teacher} \textbf{\bibinfo{volume}{45}},
  \bibinfo{pages}{260} (\bibinfo{year}{2007}).

\end{thebibliography}

\end{document}